\documentclass[prb,twocolumn,showpacs,preprintnumbers,amsmath,amssymb]{revtex4}
\usepackage{graphicx}
\usepackage{dcolumn}
\usepackage{bm}
\newcommand{\bra}[1]{\left\langle #1\right|}
\newcommand{\ket}[1]{\left| #1\right\rangle}

\begin{document}

\title{Incompressible Liquid, Stripes and Bubbles in rapidly rotating 
Bose atoms at $\nu=1$}

\author{Hidetsugu Seki and Kazusumi Ino}

\affiliation{
Department of Pure and Applied Science, University of Tokyo,
Komaba 3-8-1, Meguro-ku, Tokyo, 153-8902, Japan}


\begin{abstract}
We numerically 
study  the system of rapidly rotating Bose atoms at the filling factor
(ratio of particle number to vortex number) $\nu=1$ with 
the dipolar interaction.
A moderate dipolar interaction stabilizes the incompressible quantum liquid at
$\nu=1$. Further addition induces a collapse of it.
The state after the collapse is a compressible state which has
phases with stripes and bubbles.
 There are two types of bubbles with a different array.
We also investigate models constructed from truncated interactions and 
the models with the three-body contact interaction.
They also have phases with stripes and bubbles.

\end{abstract}

\pacs{03.75.Lm,03.75.Kk,73.43.Nq}
\maketitle

\section{\label{sec:level1}INTRODUCTION}
Recently rapidly rotating ultracold atoms in a trap have attracted considerable
interest. Experimental studies of moderately rotating atomic Bose gases 
have shown evidence for the formation of a vortex lattice
\cite{Madison00,Abo-Shaeer01}.
Bose-Einstein condensation occurs even in the case without interaction but
it is interaction which forms a vortex lattice.
The number  which parameterizes the population of vortices 
is given by $\nu=N/N_V$ where $N$ is the number of bosons and $N_V$ 
 is the average number of vortices.  
For  $\nu > \nu^*$ where $\nu^* \sim 6$, the vortex lattice 
is stable \cite{Cooper01}. 
For the high frequency regime $\nu < \nu^*$, numerical studies have predicted 
that   the vortex lattice ought to melt  and should be  replaced by 
incompressible liquids which are closely related to fractional quantum Hall states \cite{Wilkin98,Wilkin00,Cooper01}. In that regime, $\nu$ corresponds 
to  the filling  of the Landau levels in the rotating plane.
At $\nu=1/2$, the bosonic Laughlin state\cite{Laughlin83} 
is predicted \cite{Wilkin98},
 while at $\nu=k/2$ ($k$ is an integer $ \geq 2$), states with a clustered structure emerge\cite{Cooper01}. For $k=2$ it is called 
the Pfaffian state\cite{Moore91}.  
 Its fermionic counterpart is believed to be realized in the $\nu=5/2$ plateau
\cite{Morf98}
and has been proposed as an excellent candidate for 
a quantum computational device
\cite{DasSarma05}.

In typical Bose atomic gases, the interaction is short-ranged and
 can be modeled by a contact interaction\cite{Dalfovo99}.
Recently the condensation of Bose atoms with a strong 
magnetic dipole moment has been observed\cite{Griesmaier05}.
These atoms interact nonlocally through the dipolar interaction.
The effect of the dipolar interaction on vortex lattices and incompressible 
liquids above has been studied for 
$\nu=1/2$\cite{Cooper05}, $\nu=3/2$\cite{Rezayi05} 
 and $\nu=2$\cite{Cooper06}.
For these cases, a moderate dipolar interaction stabilizes 
 the incompressible liquids.
When the dipolar interaction gets strong, however, 
these states collapse.
For the $\nu=1/2$ case, the state after the collapse is identified as a stripe state in 
both Gross-Pitaevskii mean-field theory and microscopic theory 
\cite{Cooper05}. 
As the dipolar interaction gets strong, the stripe state at $\nu=1/2$ 
eventually shows a transition to a bubble state \cite{Cooper05}.
Similar transitions are also  observed for $\nu=2$ \cite{Cooper06}.
A tendency of a spectrum to form stripes is mentioned also for $\nu=3/2$
\cite{Rezayi05}.  These stripes and bubbles are closely related to 
those of two-dimensional fermions in higher Laudau levels 
\cite{Koulakov96,Rezayi99,Haldane00}.

In this paper, we consider the finite system of Bose atoms at 
$\nu=1$ (up to $N=12$) which interact  through 
the dipolar interaction and perform an exact diagonalization study to 
examine 
whether similar stabilization and collapse are induced. 
We observe the stabilization of the Pfaffian state similar to 
those of incompressible liquids at $\nu=1/2,3/2$ 
by a moderate dipolar interaction. Further addition of the dipole moment
induces a collapse of the incompressible liquid.
To investigate  the nature of the collapse,  
we take  the periodic rectangular geometry. 
This geometry is suitable to address bulk properties
 and  transitions to other states.
For the cases we have investigated up to $N=12$, 
we observe four types of  states:  stripes, bubbles, 
intermediate states of two types of stripes, and intermediate states 
of stripes and bubbles. 
These states are compressible and therefore sensitive to geometry.
The periodic rectangular geometry is parameterized by an aspect ratio ($\leq 1$).  We study the spectral flow of low-lying levels 
due to the change of the aspect ratio in detail. 
We also use  the pair distribution function and 
the guiding-center static structure factor of ground states. 
They give a consistent picture of phases of the $\nu=1$ Bose atoms.   
For the stripe states, the direction and the number of stripes depend 
on the aspect ratio and the interaction. 
For large aspect ratios,  we observe that  the increase of 
the strength of the dipolar interaction results in the formation of bubbles.
Further addition of the dipolar interaction leads to 
a transition to another bubble state which has a different array.
For small aspect ratios, stripe states seem to be favored
 even when the strength of the dipolar interaction is strong :
we observe no evidence that a transition from stripes to bubbles occurs.

To elucidate the nature of these transitions and realized states,
we  consider  models constructed from truncated interactions.
Any two-body interaction between particles in the lowest Landau level can be
expanded in pseudopotential \cite{Haldane83} $V_m, m=0,1,2,3,\cdots$, each
term of which corresponds to the projection to the two particle states with
definite relative angular momentum. For bosons, the odd terms vanish 
$V_m=0, m=1,3,5,\cdots$. For the dipolar interaction, all the even terms 
$V_m, m=0,2,4, \cdots$ appear. 
We investigate the model only with the $V_2$ interaction (``hollow-core'').
It turns out that this model 
is always in a bubble phase for the aspect ratio above $\sim 0.5$.  
A certain amount of the two-body contact interaction $V_0$ 
causes a transition to another bubble state with a  different array.
For  small aspect ratios below $ \sim 0.5$,  we observe the formation 
of stripes.

These results are for two-body interactions.  
We also address the same issue to  the three-body contact interaction, 
for which the Pfaffian state 
is the exact ground state \cite{Greiter92}.
We investigate the effect of the $V_2$ interaction on the model. 
Similar transitions to stripes and bubbles are observed also for this model.

The organization of this paper is as follows. 
In Sec. \ref{sec:level2},  
we give details of the formalism used to treat the bose problem 
on the periodic rectangular geometry.  
In Sec. \ref{sec:level3}, we show  results 
for the dipolar model.  In Sec.\ref{sec:level5} 
is devoted to the $V_2$ model.
In Sec.\ref{sec:level4}, we show  results
for the three-body contact interaction. 
Sec.\ref{sec:level6} gives conclusions.

\section{\label{sec:level2} Formalism}

The Hamiltonian describing $N$ bosons of mass $m$ in a rotating coordinate
is
\begin{eqnarray}
H&=& \sum_{i=1}^{N}\frac{1}{2m}(\mathbf{p}_i-m\omega 
\hat{\mathbf{z}}\times \mathbf{r}_i)^2 \nonumber \\
 & & + \sum_{i=1}^{N}\frac{1}{2}m(\omega_0^2-\omega^2)(x_i^2+y_i^2)
+\sum_{i=1}^{N}\frac{1}{2}m\omega_z^2 z_i^2 \nonumber \\
& & + \sum_{1\le i < j \le N}V(\mathbf{r}_i-\mathbf{r}_j),
\end{eqnarray}
where the angular velocity is $\omega \hat{\mathbf{z}}$, the $x$-$y$ trap
frequency is $\omega_0$, and the axial trap frequency is $\omega_z$.
The confinement length in the rotating plane (the $x$-$y$ plane) is 
$l \equiv \sqrt{\hbar/(m \omega_0)}$ and the confinement length along the 
$z$-direction is $l_z \equiv \sqrt{\hbar/(m \omega_z)}$. 
When $\omega_z$ is sufficiently large, the strong confinement 
along the $z$-direction freezes the motion along the $z$-direction in 
the ground state of the harmonic oscillator and the motion becomes quasi two-dimensional. 
For $\omega \sim \omega_0$, the system is equivalent to the bosons of charge
$q$ in a magnetic field $\mathbf{B}=(2 m \omega/q) \hat{\mathbf{z}}$. 
When $\omega$ is sufficiently large and interaction is weak, 
we can treat the $x$-$y$ motion to be restricted to the lowest Landau level.
For $N$ bosons spread over an area $A$, the characteristic parameter is 
$\nu\equiv N/N_V$ where $N_V=(m \omega A)/\pi \hbar$ is the average number
of vortices. In this paper, we focus our attention at $\nu=1$.

Let us next describe interactions between the atoms.
The two-body contact interaction\cite{Dalfovo99}  is given by 
\begin{equation}
V_0=g \sum_{1 \le i < j \le N}
\delta^3 (\mathbf{r}_i-\mathbf{r}_j),
\end{equation}
where $g=4\pi \hbar^2 a_s/m$ ($a_s$ is the $s$-wave scattering length).
In typical Bose atoms, the interaction between particles is short-ranged
and can be approximated excellently by $V_0$.
Next, the electric or magnetic dipole interaction is given by 
\begin{equation}
V_{\mathrm{dip}}=C_d \sum_{1 \le i < j \le N}
\frac{\mathbf{p}_i \cdot \mathbf{p}_j-
3(\mathbf{n}_{ij}\cdot \mathbf{p}_i)(\mathbf{n}_{ij}\cdot \mathbf{p}_j)}
{|\mathbf{r}_i-\mathbf{r}_j|^3},
\end{equation}
where 
$\mathbf{n}_{ij}=(\mathbf{r}_i-\mathbf{r}_j)/|\mathbf{r}_i-\mathbf{r}_j|$,
the $\mathbf{p}_i's$ are unit vectors which represent direction of 
the dipole moment. We assume that the dipole moments are parallel to
the $z$-direction. The model with  the dipolar interaction  
$H = V_0 + V_{\mathrm{dip}}$ will be called as the dipolar model below. 
We expand $H$ in pseudopotential \cite{Haldane83} 
$V_m,m=0,1,2,\cdots$.  For bosons, only even $m$ contribute. $V_m$ depends on
the trap asymmetry $l_z/l$. We assume that $\omega_z$ is sufficiently
large compared to $\omega$ so that we can take the limit that the thickness 
along the $z$-direction of the two-dimensional motion (in the $xy$ plane) is 
negligible, $l_z/l \to 0$. 
To first order in $l_z/l$, the pseudopotential is given by\cite{Cooper05}
\begin{equation}
V_0=\sqrt{\frac{2}{\pi}}\frac{\hbar^2 a_s}{m l^2 l_z}
+\sqrt{\frac{2}{\pi}}\frac{C_d}{l^2 l_z}-\sqrt{\frac{\pi}{2}}\frac{C_d}{l^3}
\end{equation}
\begin{equation}
V_{m>0}=\sqrt{\frac{\pi}{2}}\frac{(2m-3)!!}{m! 2^m}\frac{C_d}{l^3}
\end{equation}
 The strength of the dipolar interaction in the dipolar model $V_0+V_{\mathrm{dip}}$
 is controlled by $ \alpha \equiv V_2/V_0 $ .
We will also denote the potential with $V_{k} \neq 0$  for $k=i$, $V_k=0$ for $k \neq i$  as $V_i$ below.

For realistic atoms, there are higher order terms 
of  many-body interactions. The first of them is 
the  three-body contact interaction
\begin{equation}
V_{\mathrm{3b}}=C_{3b} \sum_{1\le i < j < k \le N}
\delta^3 (\mathbf{r}_i-\mathbf{r}_j) \delta^3 (\mathbf{r}_i-\mathbf{r}_k).
\end{equation}
The exact zero-energy ground state of $V_{\mathrm{3b}}$ is the Pfaffian 
state \cite{Greiter92} : 
for the disk geometry in the symmetric gauge, it is given by  
(except for the gaussian factor of the lowest Landau level) 
\begin{equation}
\Psi_{\rm Pf} ={\rm Pf}\left( \frac{1}{Z_i-Z_j}\right) 
\prod_{i <j} (Z_i-Z_j)  
\end{equation}
where $Z_i=x_i+iy_i$.
This state is expected to have quasiparticles and quasiholes with 
nonabelian statistics \cite{Moore91}. 
The modular $S$ matrices underlying the
statistics  have been studied in Ref. \cite{Ino98}.

We consider several models with these interactions 
on  the periodic rectangular  geometry. 
We denote the sides of  the rectangular as $a$ and $b$. 
The translational invariance of the system gives 
 a conserved pseudomomentum \cite{Haldane85} $\mathbf{K}=(K_x,K_y)$.
 The pseudomomentum runs over a Brillouin zone containing $\bar{N}^2$ points,
 where $\bar{N}$ is the greatest common division of $N$ and $N_V$.
$K_x,K_y$ is measured in units of $2\pi \hbar/a$ and $2 \pi \hbar/b$ 
respectively. The states at $(\pm K_x,\pm K_y)$ are degenerate 
by symmetry so we may take only positive $K_x,K_y$.

Next we recall the definition of  some functions.
The guiding-center 
 static structure factor $S_0(\mathbf{q})$ 
for a state $\ket{\Psi}$ is given by   
\begin{equation}
S_0(\mathbf{q}) = \sum_{i,j} \bra{\Psi} e^{i\mathbf{q}\cdot(\mathbf{R}_i-\mathbf{R}_j)} \ket{\Psi},
\end{equation}
where $\mathbf{R}_i$ is the guiding center of the $i$-th 
particle(See Ref. \cite{Prange90} for detail).
Also, the pair distribution
 function 
in the real space $G(\mathbf{r})$ is defined by  
\begin{equation}
G(\mathbf{r})=\frac{ab}{N(N-1)}\langle \Psi|  \sum_{i \ne j} \delta(\mathbf{r}-\mathbf{r}_i-\mathbf{r}_j)|\Psi \rangle 
\end{equation}
$G(\mathbf{r})$ and $S_0(\mathbf{q})$ 
will be used to study the nature of ground states.

\section{\label{sec:level3} RESULTS FOR THE DIPOLAR MODEL}

In this section, we present results of exact diagonalizations
 of the dipolar model $V_0+ V_{\mathrm{dip}}$ at $\nu=1$. 
The strength of the dipolar interaction is controlled by $\alpha$ 
defined in the previous section. 
As discussed in Ref.\cite{Cooper01},  the ground state at $\alpha=0$ 
 is an incompressible liquid and  has a large overlap with the Pfaffian state.   
We add a moderate dipolar interaction 
(small $\alpha$) and  examine the stability of the 
incompressible liquid.

We first fix the aspect ratio to be $1.0$.
For even $N$ up to $12$,  we observe that three nearly degenerate states 
with lowest energies appear at $\mathbf{K}$ where the Pfaffian states exist
and are clearly separated from the rest of the spectrum.

\begin{figure}
\rotatebox{-90}{
\includegraphics[width=5cm]{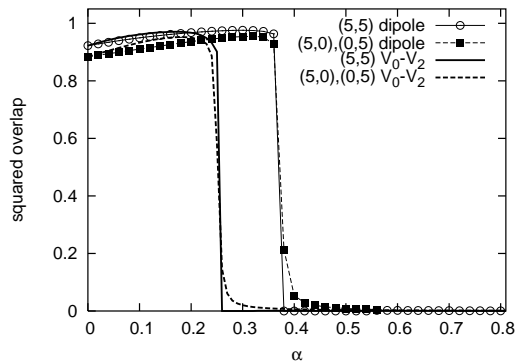}}
\caption{\label{fig1} The squared overlaps of the two degenerate 
ground states and a nearly degenerate state
for both the dipolar and  the $V_0-V_2$ model
with the Pfaffian states are shown for $N=10$ at $a/b=1.0$. 
The Pfaffian states exist at $\mathbf K=(5,5),(5,0),(0,5)$.
The two degenerate ground states appear at $\mathbf K=(5,0),(0,5)$.
The degeneracy is
due to geometrical symmetry at $a/b=1.0$.
}
\end{figure}

\begin{figure}
\begin{tabular}{c}
\rotatebox{-90}{
\includegraphics[width=4.5cm]{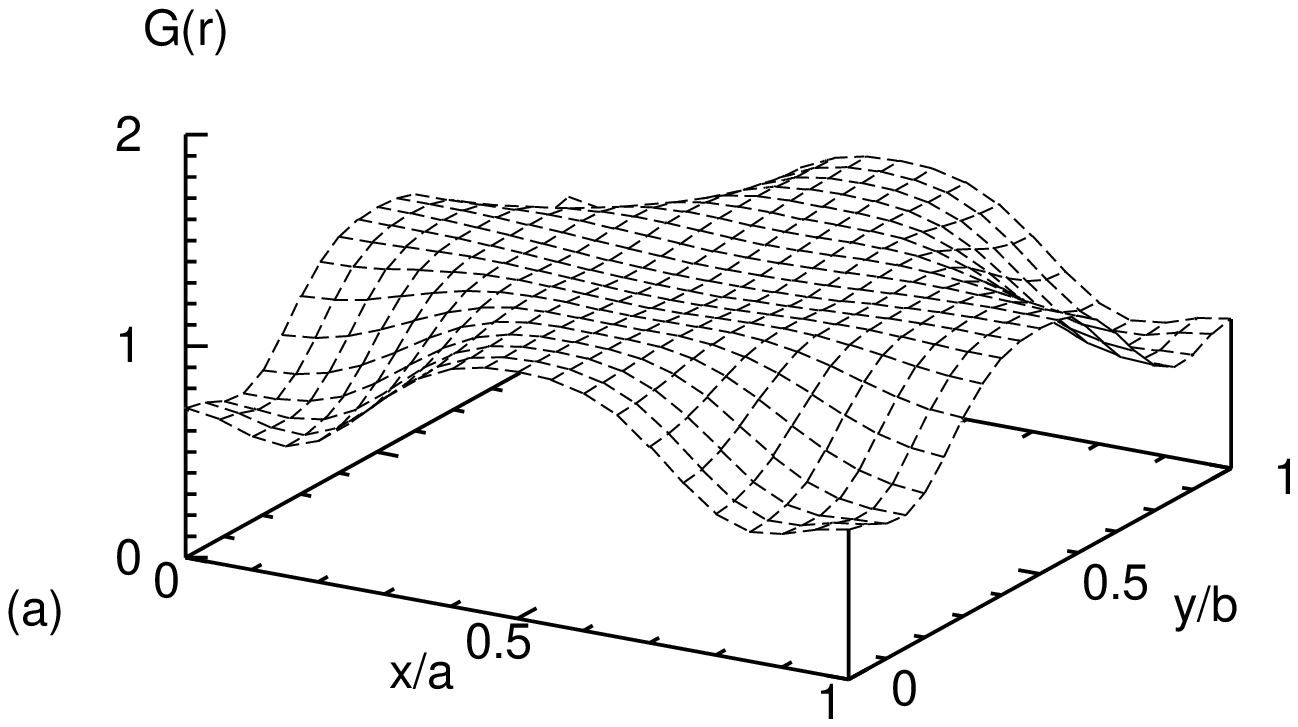}} \\
\rotatebox{-90}{
\includegraphics[width=4.5cm]{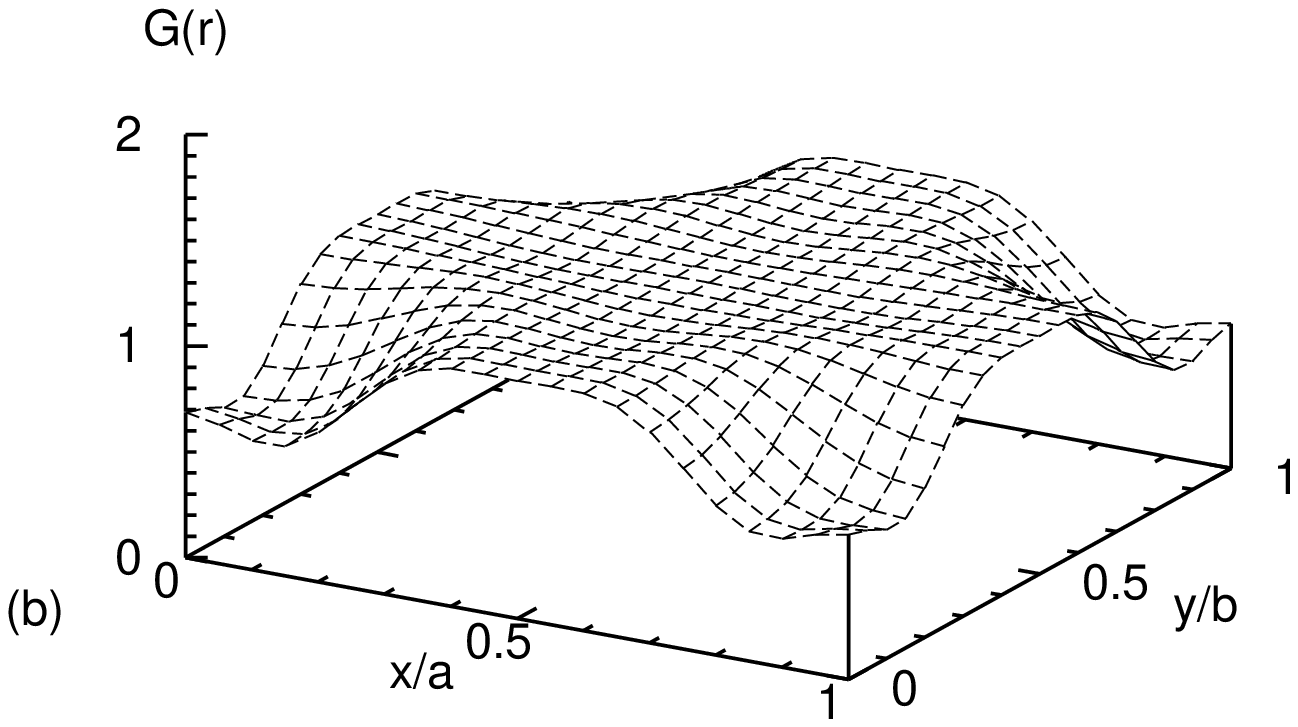}}
\end{tabular}
\caption{\label{fig2}
Pair distribution function $G(\mathbf{r})$ of various states at $a/b=1.0$.
(a) The Pfaffian state for $N=10$ at $\mathbf{K}=(0,5)$. 
(b) The ground state of the dipolar model for 
$N=12$ at $\alpha=0.30,\mathbf{K}=(6,6)$.}
\end{figure}

In Fig. \ref{fig1}, we present the squared overlaps of  
 two degenerate ground states and a nearly degenerate state 
with the Pfaffian states for $N=10$ as functions of $\alpha$. 
At small $\alpha$, the overlaps increase monotonically as $\alpha$ increases. 
The maximum of the overlaps is reached around $\alpha \sim 0.3$.
Thus a moderate dipolar interaction with $ \alpha $ below  $0.3$
stabilizes the Pfaffian states.  
Figure \ref{fig2} shows  the pair distribution function $G(\mathbf{r})$ 
of the Pfaffian state for $N=10$ and the ground state of the dipolar model
 for $N=12$ at $\alpha=0.3$. 
Both states have very similar shapes of $G(\mathbf{r})$. 
Except for regions
near four corners on the rectangular boundary, $G(\mathbf{r})$ is almost
constant and isotropic. This property indicates incompressibility of the 
Pfaffian state. 
The $G(\mathbf{r})$ of the ground state of the dipole model 
for $N=12$ is slightly  flatter than that of the Pfaffian state for $N=10$.
The nearly degenerate states in two other sectors of $\mathbf{K}$    
have similar $G(\mathbf{r})$.

When the dipolar interaction gets strong, the overlaps show 
a very abrupt drop.  
 It occurs around $\alpha \sim 0.4 $ for all the sectors  
as shown in Fig. \ref{fig1}.
We observe similar stabilization 
and collapse occur in the region $a/b = 0.3 \sim  1.0$. 
The strength of the dipolar interaction $\alpha^{*}$ where the 
collapse occurs vary with the aspect ratio.
If we decrease the aspect ratio, $\alpha^{*}$ gradually increases and 
 reaches to $\sim 0.5$ at $a/b=0.3$. We also confirm that similar
stabilization and collapse occur for $N=6,8$ at $\alpha = 0.4 \sim 0.5$
for $a/b =0.3 \sim  1.0$.

\begin{figure}
\begin{center}
\rotatebox{-90}{
\includegraphics[width=5cm]{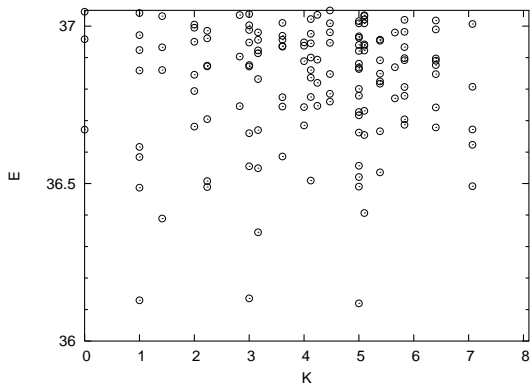}}
\caption{\label{fig3} 
Energy spectrum of the dipolar model 
for $N=10$ at $\alpha=0.65,a/b=0.8$ 
as a function of $K=\sqrt{K_x^2+K_y^2}$.
The ground state at $\mathbf{K}=(0,5)$ and states at $\mathbf{K}=(0,3),(0,1)$
are quasidegenerate.}
\end{center}
\end{figure}

We next investigate states after the collapse of the incompressible liquid. 
In Fig. \ref{fig3}, we present the energy spectrum for $N=10$ at 
$\alpha=0.65$ and $a/b=0.8$. 
The ground state at $\mathbf{K}=(0,5)$ and the lowest energy states at 
$\mathbf{K}=(0,3),(0,1)$ are quasidegenerate and clearly separated from the 
rest of the spectrum. A characteristic wave vector that separates these
quasidegenerate states is $\mathbf{K}^*=(0,2)$. This suggests that 
these quasidegenerate ground states have 
a tendency to form a stripe state. 

\begin{figure}
\begin{center}
\begin{tabular}{c}
\rotatebox{-90}{
\includegraphics[width=4.5cm]{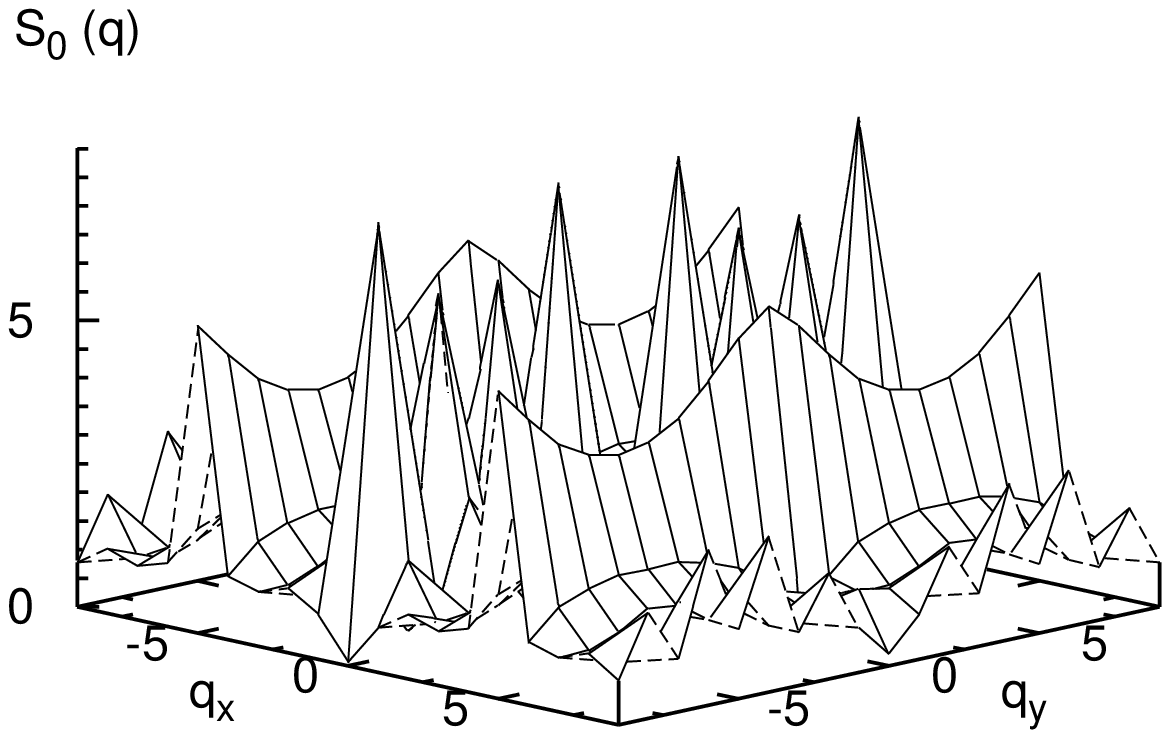}} \\
\rotatebox{-90}{
\includegraphics[width=4.5cm]{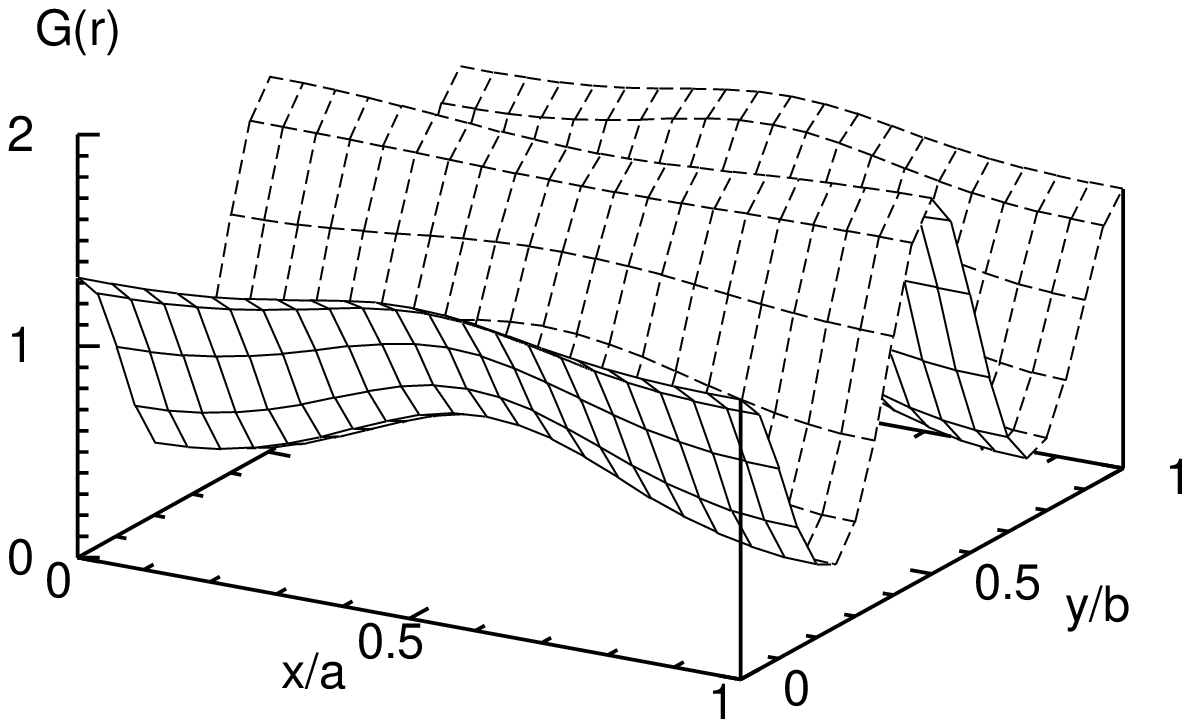}}
\end{tabular}
\caption{\label{fig4} 
Guiding-center static structure factor
$S_0 (\mathbf{q})$ and pair distribution function $G(\mathbf{r})$ 
of the ground state of the dipolar model
for $N=10$ at $\alpha=0.65,a/b=0.8,\mathbf{K}=(0,5)$.
}
\end{center}
\end{figure}

Let $N_D$ be the number of distinct quasidegenerate ground states,
and let $N_{s}$ be the number of stripe.
There are $\bar{N}^2$ distinct values in the Brillouin zone, 
where $\bar{N}$ is the greatest common divisor of $N$ and $N_V$.
If the translational symmetry is broken in one direction, 
the length of the Brillouin zone of the one-dimensional superlattice is 
$(2\pi \bar{N})/(LN_D)$ 
where $L$ is the period of the superlattice. 
It must be $2\pi/(L/N_s)$ where $L/N_s$ 
is the length per stripe. This yields $N_sN_D=\bar{N}$.
For $\nu=1$,$\bar{N}=N$. The relation $N_{s}N_D=N$ yields  
the number of boson per stripe  $M = N/N_{s}=N_D$.
In the case above,  $N=10$ and $N_D=5$, 
which gives $N_{s}=2$ and $M=5$.

To study properties of these
quasidegenerate states, we calculate the guiding-center 
 static structure factor $S_0 (\mathbf{q})$ and the pair distribution 
function $G(\mathbf{r})$ of the ground state. We measure $q_x$ and $q_y$
in units of $2\pi \hbar/a$ and $2\pi \hbar/b$.
In Fig. \ref{fig4}, we present $S_0 (\mathbf{q})$ of the ground state 
for $N=10$ at  $\alpha=0.65$, $a/b=0.8,\mathbf{K}=(0,5)$.
There are peaks at $\pm \mathbf{q}^*=(0,\pm 2)$ and $\pm 2\mathbf{q}^*$. 
Other peaks at $\pm 3\mathbf{q}^*, \pm4\mathbf{q}^*$
are due to the translational symmetry at $\nu =1$.
The peaks at $\pm \mathbf{q}^*=(0,\pm2)$ indicate a strong density-density
correlation in the ground state at this ordering vector.
In Fig. \ref{fig4}, we also present $G(\mathbf{r})$ of the state. 
The shape of $G(\mathbf{r})$ is quite different from that of  
the Pfaffian state shown in Fig. \ref{fig2}.
$G(\mathbf{r})$ depends on the $y$-direction but 
has little dependence along the $x$-direction.
 Two peaks are seen along the $y$-direction in the unit cell, 
which is consistent with  the peak of $S_0 (\mathbf{q})$ at 
$\mathbf{q}^*=(0,2)$.
 $S_0 (\mathbf{q})$ and $G(\mathbf{r})$ indicate that 
this ground state is a stripe state with two stripes lying parallel to 
the $x$-direction. This is consistent with the analysis of 
$N_{\mathrm{s}}$ given in the previous paragraph.
 $S_0 (\mathbf{q})$ and $G(\mathbf{r})$ of other quasidegenerate states at 
$\mathbf{K}=(0,3),(0,1)$ are similar to those of the ground state
at $\mathbf{K}=(0,5)$.

\begin{figure}
\begin{center}
\rotatebox{-90}{
\includegraphics[width=5cm]{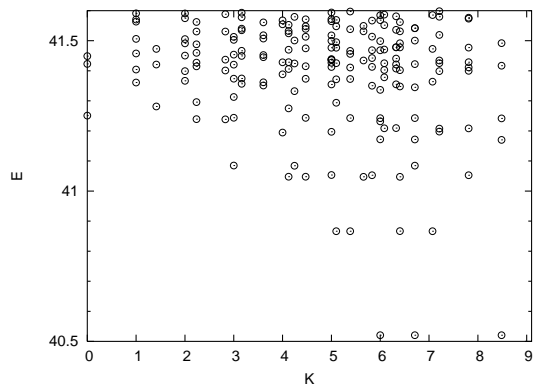}}
\caption{\label{fig5} 
Energy spectrum of the dipolar model 
for $N=12$ at $\alpha=0.6,a/b=0.5$.
The ground state at $\mathbf{K}=(6,0)$ and states at $\mathbf{K}=(6,3),(6,6)$
are quasidegenerate.
The lowest energy states at $\mathbf{K}=(5,1),(5,2),(5,4),(5,5)$ 
form a quasidegenerate first-excited states.}
\end{center}
\end{figure}

\begin{figure}
\begin{center}
\begin{tabular}{c}
\rotatebox{-90}{
\includegraphics[width=4.5cm]{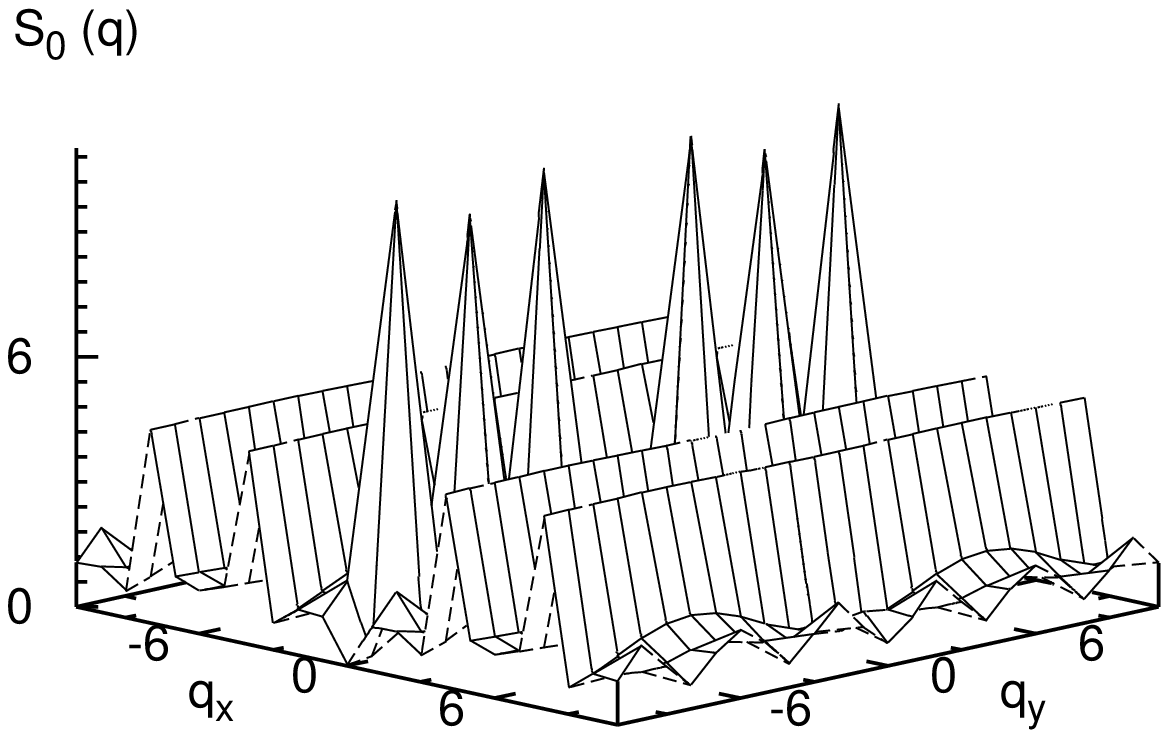}} \\
\rotatebox{-90}{
\includegraphics[width=4.5cm]{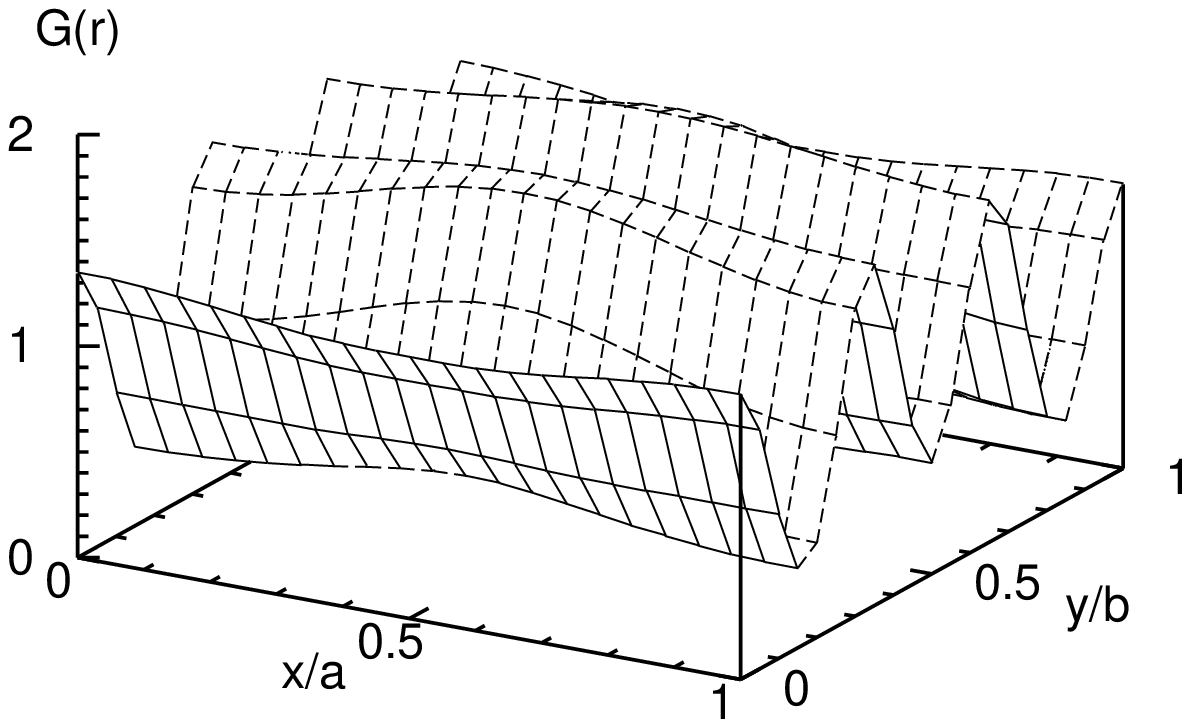}}
\end{tabular}
\caption{\label{fig6} 
$S_0 (\mathbf{q})$ and $G(\mathbf{r})$ of the ground state of the 
dipolar model 
for $N=12$ at $\alpha=0.6,a/b=0.5,\mathbf{K}=(6,0)$.
}
\end{center}
\end{figure}

In Fig. \ref{fig5}, we present the energy spectrum for $N=12$ at 
$\alpha=0.6$, $a/b=0.5$. 
The ground state at $\mathbf{K}=(6,0)$ and the lowest energy state at 
$\mathbf{K}=(6,3),(6,6)$ are quasidegenerate and clearly separated from the
rest of the spectrum. These states are separated
by a characteristic wave vector $\mathbf{K}^*=(0,3)$.  
The direction of stripes is parallel to the $x$-direction and the number
of stripes is three.
Fig. \ref{fig6} shows $S_0 (\mathbf{q})$ and $G(\mathbf{r})$ of the ground 
state.  There are peaks at $\pm{\mathbf{q}^*}=(0,\pm3)$ and at 
$\pm 2 {\mathbf{q}^*} $. Other peaks at $\pm 3 {\mathbf{q}^*} $ are
due to the translational symmetry. 
We also present $G(\mathbf{r})$ of the ground
state in Fig. \ref{fig6}. There are three peaks along the $y$-direction.
This is consistent with the peak of $S_0 (\mathbf{q})$ at $\mathbf{q}^*$.
Quasidegenerate states at $\mathbf{K}=(6,3),(6,6)$ show similar 
$S_0 (\mathbf{q})$ and $G(\mathbf{r})$. 
In Fig. \ref{fig5}, we also observe  a clear presence of 
 nearly degenerate first-excited states formed by 
the lowest energy states at $\mathbf{K}=(5,1),(5,2),(5,4),(5,5)$. 
They are created from the quasidegenerate ground states 
by vectors $\mathbf{e}_1=(1,1),\mathbf{e}_2=(-1,1)$.  
A part of a less clear array of  
nearly degenerate second-excited states in Fig. \ref{fig5} 
is also generated from the first-excited states by these vectors.
Some other excited states in the second-excited array have the quantum numbers 
generated from $(1,0)(-1,0)$.  
The appearance of clearly separated  low-energy   bands 
is accounted for  by 
 low-energy particle-hole excitations in which a boson 
is moved from one stripe to another. These bands are almost flat, 
showing that such a hopping of the bosons 
between stripes is strongly suppressed.
This band structure gives evidence of the stripe state.
The band for excited states is not clear in Fig. \ref{fig3} for $N=10$.
For $N=12$, the stripe state  with three stripes appears in the region
$a/b=0.4 \sim 0.65$. 
At $a/b=0.5$, we confirm that these states persist
 for $\alpha =0.5 \sim  0.8$. 
Similarly, we observe 
quasidegenerate ground states forming two stripes lying parallel to
the $y$-direction in the region $a/b =0.7 \sim  1.0$. 
For $\alpha$ beyond $0.6$ at $a/b =0.7 \sim  1.0$, 
quasidegenerate ground states persist to be present 
 but their ordering is different from stripes (see below).

\begin{figure}
\rotatebox{-90}{
\includegraphics[width=5cm]{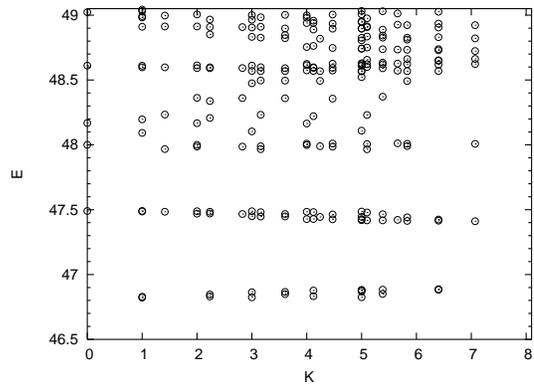}}
\caption{\label{fig7} Energy spectrum of the dipolar model for 
$N=10$ at $\alpha=1.0,a/b=0.8$. 
The quasidegenerate states are 
at $\mathbf{K}=$(1,0),(0,1),(2,1),(1,2)(3,0),(0,3),(2,3),(3,2)
,(4,1),(1,4),(4,3),(3,4),(5,0),(0,5),(4,5),(5,4),(2,5),(5,2).} 
\end{figure}

\begin{figure}
\begin{center}
\rotatebox{-90}{
\includegraphics[width=4.5cm]{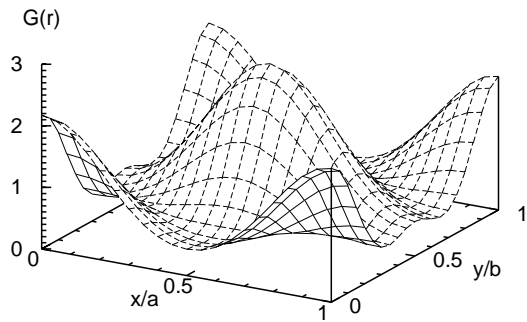}}
\end{center}
\caption{\label{fig8} $G(\mathbf{r})$ of 
the ground state of the dipolar model
 for $N=10$ at $\alpha=1.0,a/b=0.8,\mathbf{K}=(0,5)$.}
\end{figure}

Let us next turn to the region of larger dipole moment.   In Fig. \ref{fig7},
we present the energy spectrum for $N=10$ at $\alpha=1.0$, $a/b=0.8$. 
There are several quasidegenerate states whose locations in the reciprocal
space form   a lattice. This means that these states have a tendency
to form a bubble state. The primitive vectors are $\mathbf{e}_1$ and 
$\mathbf{e}_2$. 
Let $N_b$   be the number of bubbles per the unit cell.
As derived in Ref.\cite{Haldane00}, 
the relation $N_bN_D=N^2$   
gives the number of bosons in a bubble   
$M = N/N_b=N_D/N$.  
In this case, $N=10$ and $N_D=50$, which gives  
$N_b=2$ and $M=5$.  
In Fig. \ref{fig7}, one also sees quasidegenerate first-excited levels 
above the quasidegenerate ground states.  
A part of their quantum numbers are generated by 
 vectors $(\pm 1,0), (0,\pm 1)$  
from those of the quasidegenerate ground states. 
The rest of them is the first-excited state at $\mathbf{K}$ where 
the quasidegenerate ground states appear.
These almost flat bands are accounted for by 
 low-energy particle-hole excitations in which a boson 
is moved from one bubble to another.
In Fig. \ref{fig8}, we show the pair distribution function 
$G(\mathbf{r})$  at $\mathbf{K}=(0,5)$. 
There are two peaks at $\mathbf{r}=(0,0)$ and $\mathbf{r}=(a/2,b/2)$, 
which is consistent with the number of $N_b$.
$G(\mathbf{r})$ for other quasidegenerate states is similar. 
These give evidence that  these quasidegenerate states form a bubble state. 
The transition from the stripe state to the bubble state
 is observed around $\alpha \sim 0.8$.
For $N=8$, bubble states  
are seen in the region $\alpha = 0.8 \sim  2.0$, $a/b =0.4 \sim 1.0$.

\begin{figure}
\rotatebox{-90}{
\includegraphics[width=5cm]{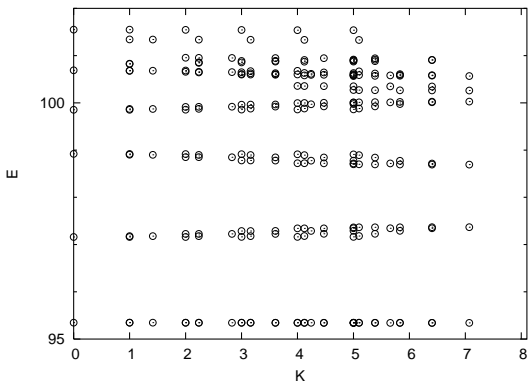}}
\caption{\label{fig9} Energy spectrum of the dipolar model for 
$N=10$ at $\alpha=3.0,a/b=0.8$. 
Every lowest energy state in the reciprocal space
is quasidegenerate.} 
\end{figure}

\begin{figure}
\begin{center}
\rotatebox{-90}{
\includegraphics[width=4.5cm]{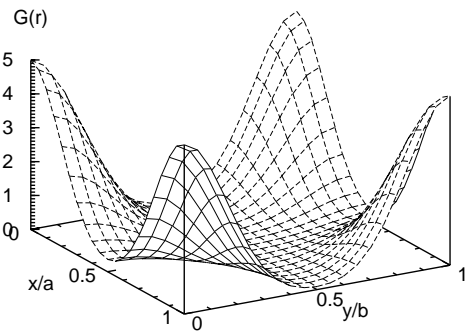}}
\end{center}
\caption{\label{fig10} $G(\mathbf{r})$ of the ground state of the 
dipolar model 
for $N=10$ at $\alpha=3.0,a/b=0.8,\mathbf{K}=(0,5)$.}
\end{figure}

In Fig. \ref{fig9}, we present the energy
spectrum for $N=10$ at $\alpha=3.0$,  $a/b=0.8$. 
The number of the quasidegenerate ground states increases compared to 
the $\alpha=1.0$ case.
The lowest states for every $\mathbf{K}$  become  quasidegenerate.
 A clear presence of the band structure in Fig. \ref{fig9} 
is readily interpreted by low-energy particle-hole excitations.
These give evidence of a bubble state.
In this case,  $N_b=1$ and $M=10$.
In Fig. \ref{fig10},  $G(\mathbf{r})$ of 
the lowest energy state at $\mathbf{K}=(0,5)$ is presented. 
The density concentrates near the corner of the unit cell, 
which is consistent with $N_b=1$.
Thus, the bubble state observed here is 
different from the bubble state for lower $\alpha$. 
The transition between these bubble states 
occurs around  $\alpha \sim 2.0$ for $N=8,10$. 

\begin{figure}
\begin{center}
\begin{tabular}{c}
\rotatebox{-90}{
\includegraphics[width=5.8cm]{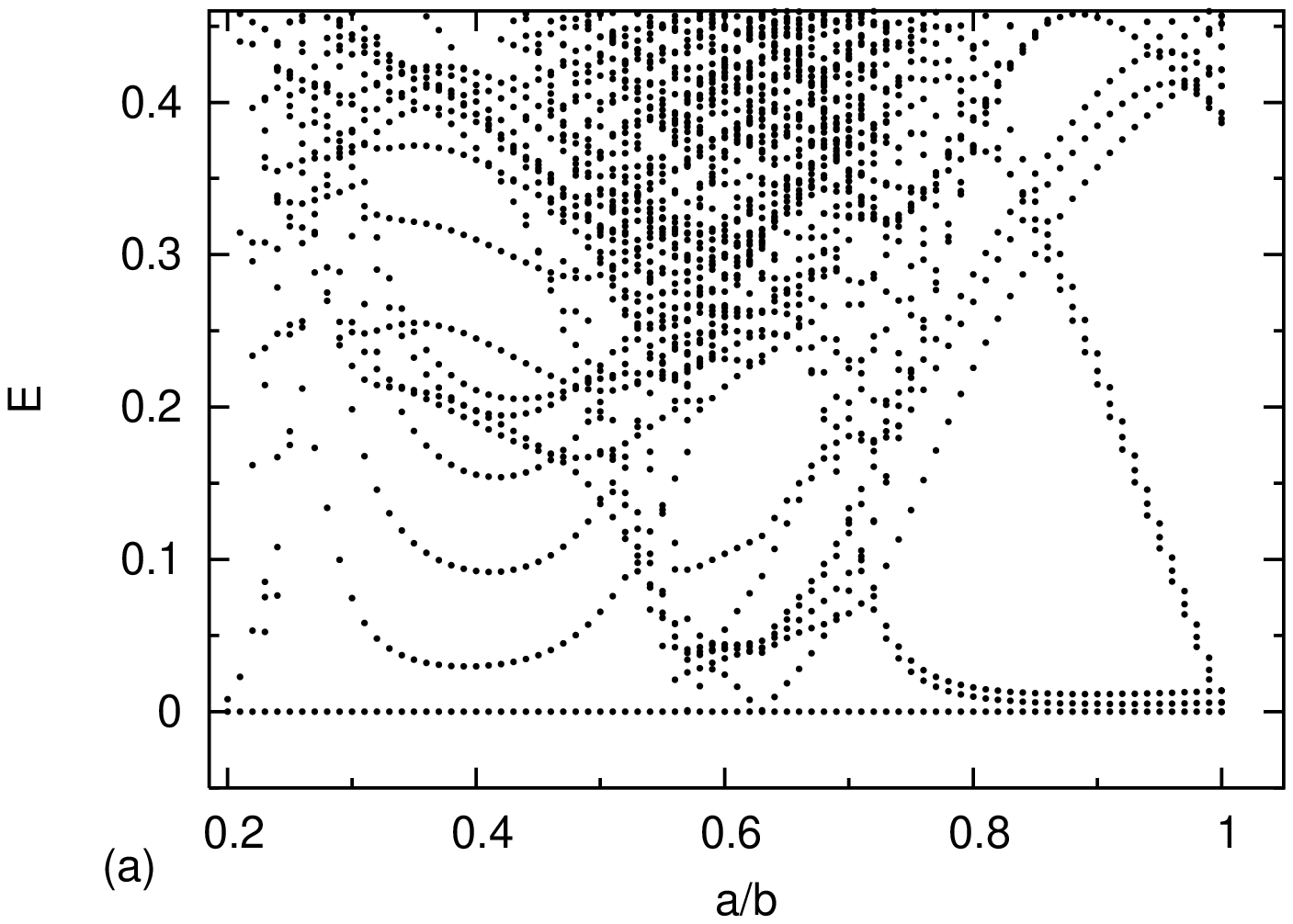}} \\
\rotatebox{-90}{
\includegraphics[width=5.8cm]{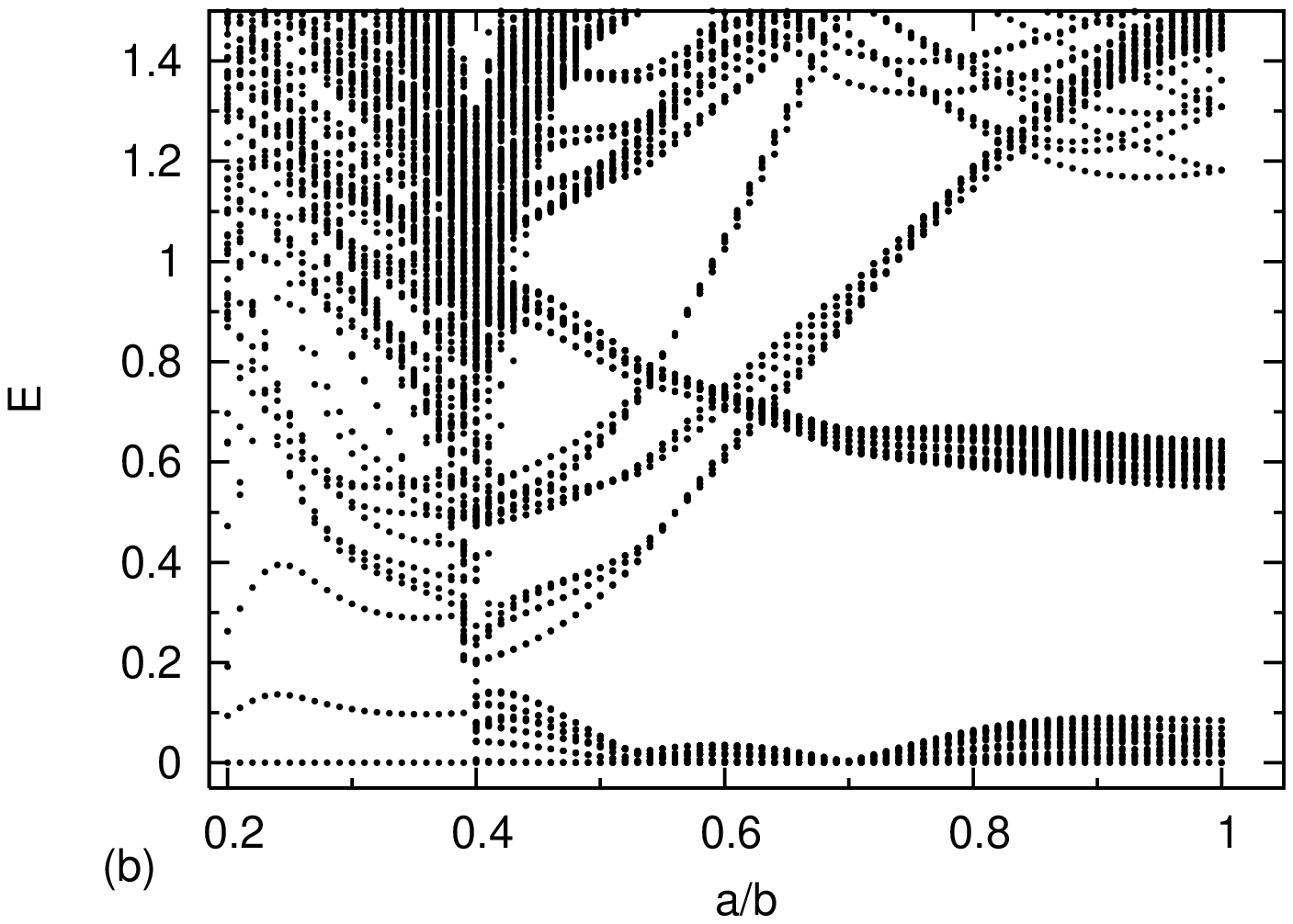}} \\
\rotatebox{-90}{
\includegraphics[width=5.8cm]{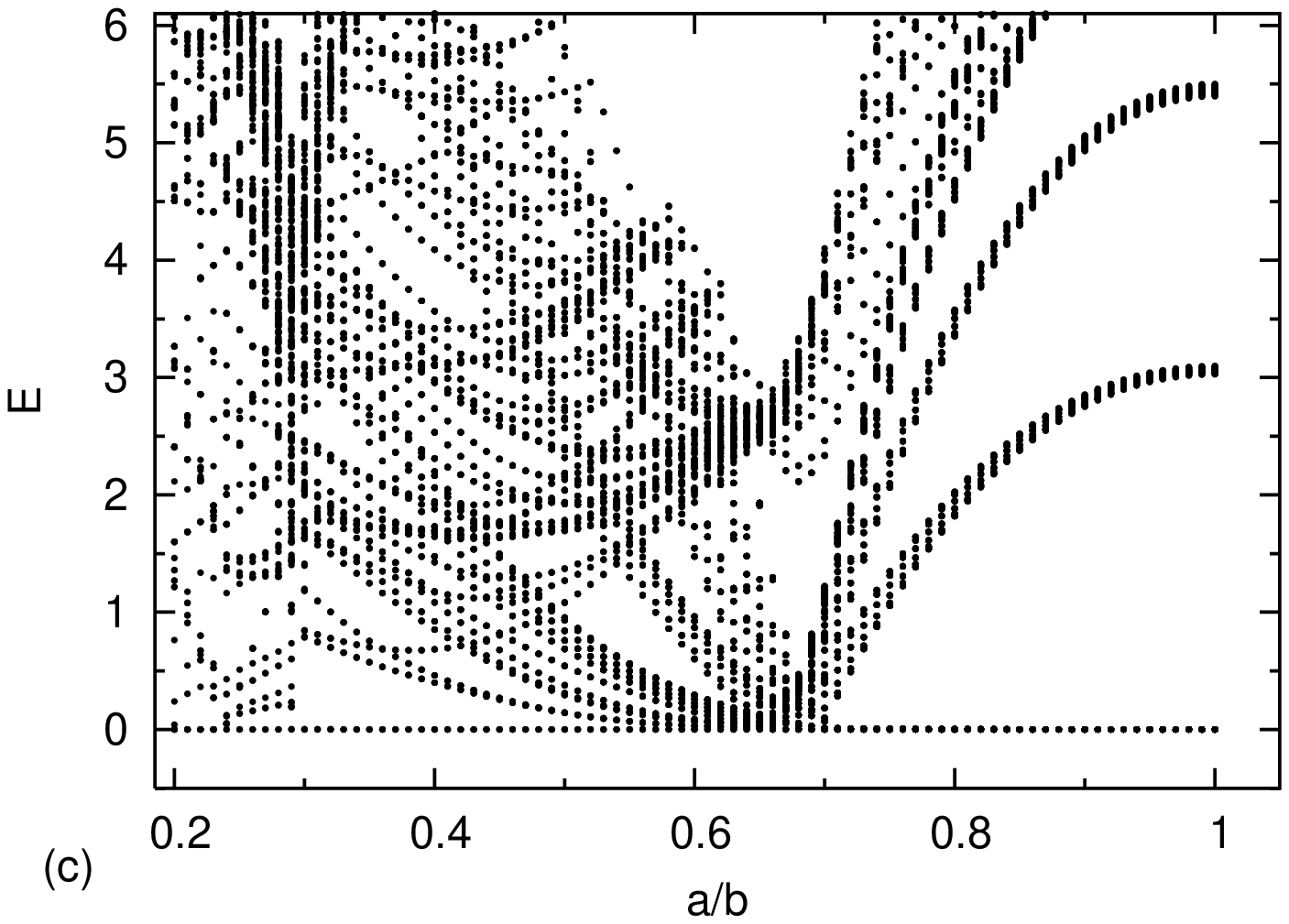}}
\end{tabular}
\caption{\label{fig11} Energy levels versus 
the aspect ratio
$a/b$ for $N=10$. 
(a)$\alpha=0.65$,
(b)$\alpha=1.0$,
(c)$\alpha=3.0$.}
\end{center}
\end{figure}

As seen above, after the collapse of the incompressible liquid, 
 ground states  are not  separated from excited 
states by a gap.  This is characteristic of a compressible state. 
In general, its property is  sensitive to geometry (the aspect ratio).  
Therefore we vary the aspect ratio for values of $\alpha$ 
and study the dependence of spectra 
on the aspect ratio.  
In Fig. \ref{fig11}, we present energy spectra 
for $N=10$ at $\alpha=0.65,1.0,3.0$. 
For  $\alpha=0.65$, changes of the level structure
accompanying level crossings are seen around $a/b \sim 0.75,0.52,0.25$. 
At $\alpha=1.0$,  a structural change is seen around $a/b \sim 0.4$. 
At $\alpha=3.0$, structural changes are seen  around $a/b \sim 0.7, 0.55,0.3$. 
Similar structural changes are observed  for other $N$ 
but the values of aspect ratios at which they occur  depend on $N$.
When structural changes are not sharp, intermediate states appear 
around  these  points as seen in Fig. \ref{fig11}.

\begin{figure}
\begin{center}
\rotatebox{-90}{
\includegraphics[width=5cm]{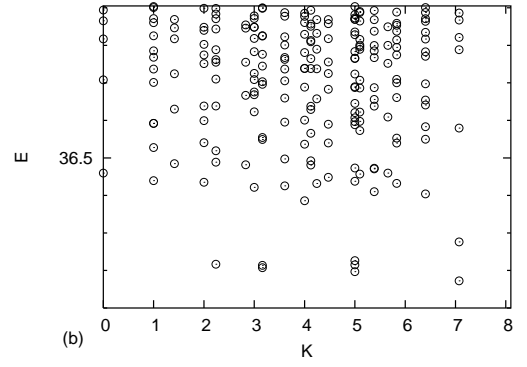}}
\caption{\label{fig12} 
Energy spectrum of the dipolar model 
for $N=10$ at $\alpha=0.65,a/b=0.6$.}
\end{center}
\end{figure}

Let us illustrate  a complicated behavior at $\alpha=0.65$  
in  the region  $a/b = 0.5 \sim  0.7$ where one sees 
 many level crossings as in Fig. \ref{fig11}(a). 
As an example, we present  the energy spectrum for $N=10$ at 
$a/b=0.6$ in Fig. \ref{fig12}. 
The ground state appears at $\mathbf{K}=(5,5)$. 
The locations of  quasidegenerate states in the reciprocal space 
do not form any lattice. 
For $N=12$, similar states appear in the region of high aspect ratios
$ a/b =0.65 \sim   1.0$ around  $\alpha = 0.65$.
We interpret them as 
 intermediate states between stripe states and bubble states 
 due to finite-size effects. 
The region  for which such intermediate states appear 
 should shrink as $N$ gets large.

Let us next turn to the highly anisotropic region ($ a/b$ below $ 0.5$) of 
Fig. \ref{fig11}(a). Energy spectrum and $G(\mathbf{r})$ shows that 
stripe states appear in the region $ a/b =0.25 \sim  0.5$.
The quasidegenerate states in this region 
 form a stripe state with three stripes lying parallel to the $x$-direction 
\cite{N=8.3}.

Let us next illustrate some details of 
 the level structure for $\alpha=1.0$ shown in Fig. \ref{fig11}(b).
The level structure of the region $ a/b =0.2 \sim  0.4$ is different 
from  those for the bubble state seen  in the region $a/b = 0.4 \sim  1.0$.
The level structure of the bubble state in the region $a/b = 0.4 \sim  1.0$
persists for $\alpha = 0.8 \sim  2.2$.
As seen in Fig.\ref{fig11}(b), 
it disappears around $a/b =0.4$ and changes to  that of a stripe state. 
 For other $\alpha$'s, 
this structural change occurs  around $a/b = 0.4 \sim  0.5$. 
 The direction of stripes is parallel to the $x$-direction. 

In Fig. \ref{fig11}(c), we present energy spectra versus the aspect ratio
for $N=10$ at $\alpha=3.0$. It shows a change of 
the level structure around $a/b \sim 0.7, 0.54$. 
The bubble states  disappear for lower aspect ratios.
Stripe states appear at $a/b$ below $0.54$. 

In stripe states in Fig. \ref{fig11}(a),(b),(c) at the highly anisotropic
region,  the direction of stripes is always 
parallel to the $x$-direction 
(the shorter axis). The number of stripes 
depends on $N$, $\alpha$, and $a/b$.

Let us next turn to the region of 
 extremely low aspect ratios 
 below  $0.2$.
In this region,  some strange behaviors are observed. 
For $N=8,10$, the overlap of  quasidegenerate ground states 
 at $\alpha=0$ with the Pfaffian states is large only for 
 one sector of $\mathbf{K}$ . 
It  drops rapidly when we move to  $\alpha \neq 0$ 
but the other two sectors for quasidegenerate ground states 
 begin to have  a large overlap with the Pfaffian states.
The collapse of these two states  
occurs at much higher value of $\alpha$ than for nonextreme aspect ratios. 
In this region, the shapes of $S_0(\mathbf{q})$ and 
$G(\mathbf{r})$ for the Pfaffian states are 
quite similar to those of stripe states.
They cannot be distinguished from
 a stripe state  with four stripes ($N=8$), five stripes ($N=10$) 
lying parallel to the $x$-direction \cite{Note2}.

Let us next present results for the isotropic case $a/b=1.0$ where 
 geometric symmetry is enhanced.
Also for this case, two stripe states appear for $N=10$ in the 
region $\alpha =0.4 \sim  0.8$.
States at $\mathbf{K}=(K_1,K_2)$ and $\mathbf{K}=(K_2,K_1)$ 
are exactly degenerate due to geometrical symmetry present at $a/b=1.0$. 
The ground states appear at $\mathbf{K}=(0,5)$ and $\mathbf{K}=(5,0)$.
States at $\mathbf{K}=(1,0),(3,0),(5,0),(0,1),(0,3),(0,5)$ are 
quasidegenerate and separated from the rest of the spectrum. 
The states at $\mathbf{K}=(1,0),(3,0),(5,0)$ form two 
stripes parallel to the $y$-direction while the states at 
$\mathbf{K}=(0,1),(0,3),(0,5)$ form two stripes parallel to the $x$-direction. 
 Thus the ground state at $a/b=1.0$ is a linear combination of these 
two stripe states.
\begin{figure}
\rotatebox{-90}{
\includegraphics[width=4.5cm]{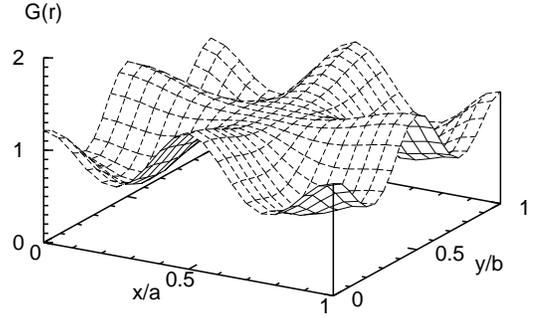}}
\caption{\label{fig13}
$G(\mathbf{r})$ of the ground state of the dipolar 
model for 
$N=12,\alpha=0.6,a/b=1.0$ at $\mathbf{K}=(6,6)$.}
\end{figure}
Similarly the ground state at $\mathbf{K}=(6,6)$ for $N=12$ 
is not a stripe state with a broken translational symmetry for one  direction, 
nor a bubble state with a broken translational symmetry for two directions.
In Fig. \ref{fig13}, we present $G(\mathbf{r})$ of the 
ground state at $\alpha=0.6,\mathbf{K}=(6,6)$.
They indicate that the ground state has a strong density-density correlation  
along two stripes lying parallel to the $x$- and $y$-directions. 
We observe that this shape of $G(\mathbf{r})$  at $a/b=1.0$ 
persists for $N=10,\alpha = 0.4 \sim 0.8$. 

\begin{figure}
\begin{center}
\begin{tabular}{c}
\rotatebox{-90}{
\includegraphics[width=4.5cm]{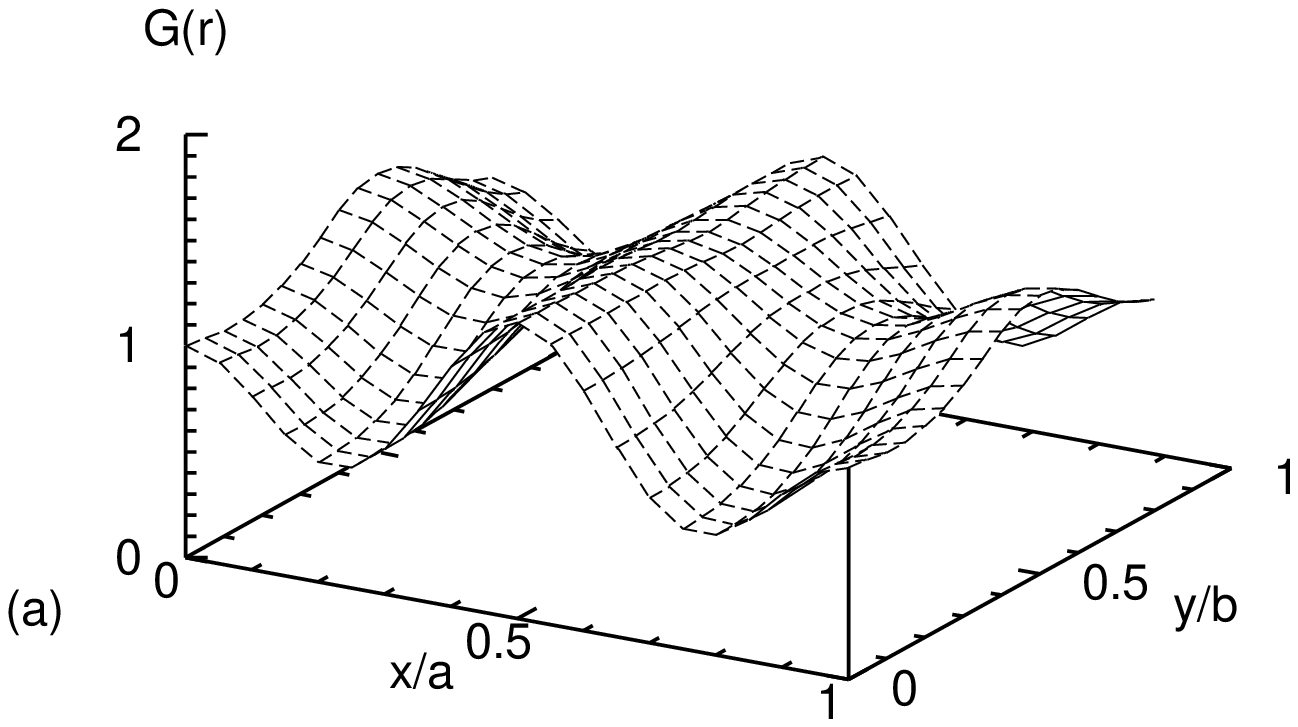}} \\
\rotatebox{-90}{
\includegraphics[width=4.5cm]{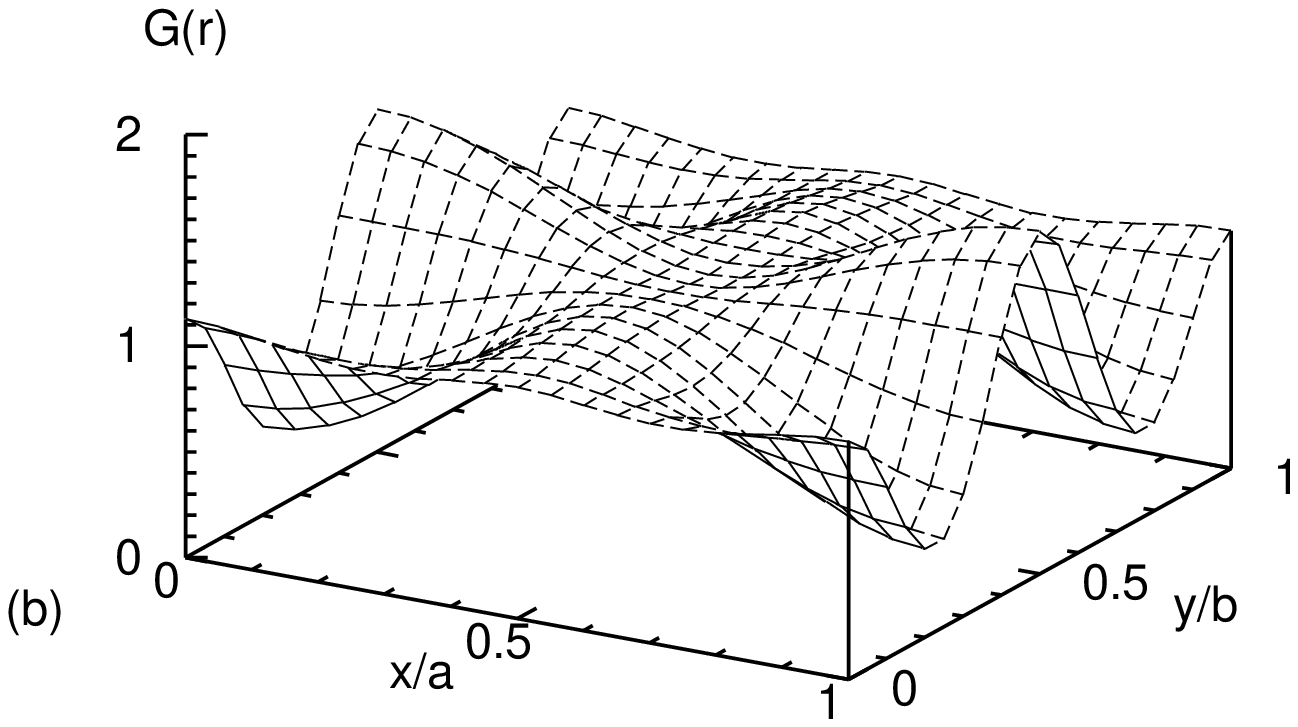}} 
\end{tabular}
\end{center}
\caption{\label{fig14} $G(\mathbf{r})$ of various
ground states of the dipolar model.
(a) $N=10,\alpha=0.49,\mathbf{K}=(5,0)$.
(b) $N=10,\alpha=0.50,\mathbf{K}=(0,5)$.}
\end{figure} 

Near the isotropic case, one expects that there may be a competence between 
 stripe states with perpendicular directions.
This is actually the case  for $N=10$ in a region around $a/b \sim 0.8$.
The direction of the stripes changes as one varies $\alpha$.
There is a level crossing at $\alpha \sim 0.496$ for $a/b=0.8$.
In the region $\alpha < 0.496$, the ground state appears at $\mathbf{K}=(5,0)$.
At $\alpha \sim 0.496$, the lowest energy states at $\mathbf{K}=(5,0)$ and 
$\mathbf{K}=(0,5)$ are quasidegenerate. In the region $\alpha > 0.496$,
the ground state appears at $\mathbf{K}=(0,5)$.
We present $G(\mathbf{r})$ of the ground states for  
$\alpha=0.49,0.50$ in Fig. \ref{fig14} (a),(b).
In the region with $\alpha < 0.496$, 
the quasidegenerate ground states at $\mathbf{K}=(5,0),(3,0),(1,0)$ 
form a stripe state with
 two stripes lying parallel to the $y$-direction.
 At $\alpha=0.50$, the ground state at $\mathbf{K}=(0,5)$ is 
an intermediate state between two stripes with perpendicular directions.  
At $\alpha=0.65$, the stripe state with 
two stripes lying parallel to the $x$-direction appears.

Before ending this section,  
let us  briefly discuss the $V_0$-$V_2$ model. 
Previously the collapse of the incompressible liquid for the two-body contact interaction 
$V_0$ by the $V_2$ interaction   
has been observed in the spherical geometry\cite{Regnault03,Regnault06}. 
In Fig. \ref{fig1}, we present the squared overlaps 
of two  degenerate ground states and a nearly degenerate state
with the Pfaffian states for $N=10$ at $a/b=1.0$.
As in the case of the dipolar interaction, a moderate $V_2$ interaction
stabilizes the Pfaffian state. It collapses
around $\alpha \sim 0.25$ as seen in Fig. \ref{fig1}. 
We confirm that quasidegenerate ground states with 
a tendency to form stripes appear after the collapse.  
When we further increase $\alpha$,  two kinds of quasidegenerate 
ground states with characteristics to form bubbles appear as 
in the case of the dipolar model.

\section{\label{sec:level5} RESULTS for the $V_2$ model}

\begin{figure}
\begin{center}
\rotatebox{-90}{
\includegraphics[width=5cm]{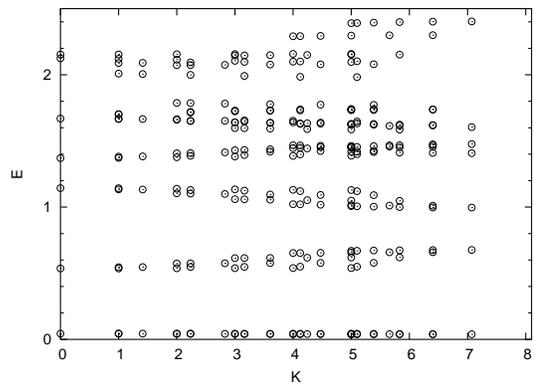}}
\caption{\label{fig15} Energy spectrum of the $V_2$ model
for $N=10$ at $a/b=0.8$.}
\end{center}
\end{figure}

In this section, we present results for the $V_2$ model.
We present the energy spectrum for $N=10$ at $a/b=0.8$ 
in Fig. \ref{fig15}.
The lowest energy states at every point in the 
reciprocal space are quasidegenerate and separated from the rest of 
the spectrum. 
Also, in Fig. \ref{fig15}, several flat excited  bands are  clearly 
seen. It is consistent with the presence of 
low-energy particle-hole excitations. 
These observations give  evidence of a tendency to
the formation of a bubble state.  
\begin{figure}
\begin{center}
\rotatebox{-90}{
\includegraphics[width=4.5cm]{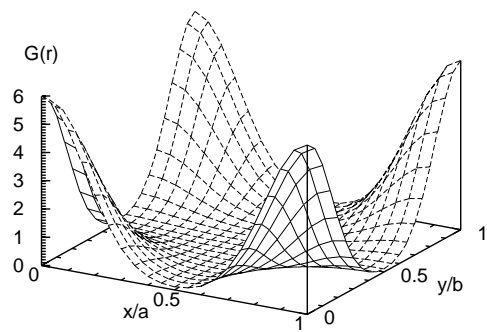}}
\end{center}
\caption{\label{fig16} $G(\mathbf{r})$ of the ground
state of the $V_2$ model for $N=12$ at $a/b=1.0,\mathbf{K}=(6,6)$.}
\end{figure} 
The pair distribution function 
$G(\mathbf{r})$ of quasidegenerate states are similar to
those of the bubble states of the dipolar mode at $\alpha=3.0$.  
Similar shapes of $G(\mathbf{r})$ are observed for 
$N=8,12$ at $a/b=0.8 \sim 1.0$. Fig. \ref{fig16} shows 
$G(\mathbf{r})$ for $N=12,a/b=1.0$.

\begin{figure}
\begin{center}
\rotatebox{-90}{
\includegraphics[width=5.8cm]{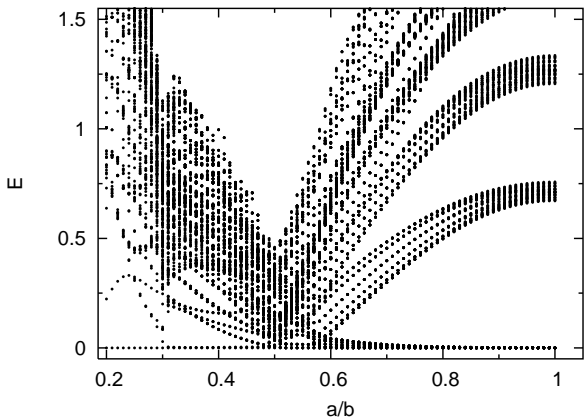}}
\caption{\label{fig17} Energy levels versus the aspect ratio
of the $V_2$ model for $N=10$.}
\end{center}
\end{figure}

We present energy spectra versus the aspect ratio for $N=10$
in Fig. \ref{fig17}.
The spectral flow of the $V_2$ model
is similar to that of the dipolar model at $\alpha$ above $2.0$.
As seen in Fig. \ref{fig17},  quasidegenerate ground states 
forming a bubble state persist for $a/b= 0.5 \sim 1.0$. 
The level structure  changes to that of  a stripe state  
around $a/b=0.45$.
\begin{figure}
\begin{center}
\rotatebox{-90}{
\includegraphics[width=5cm]{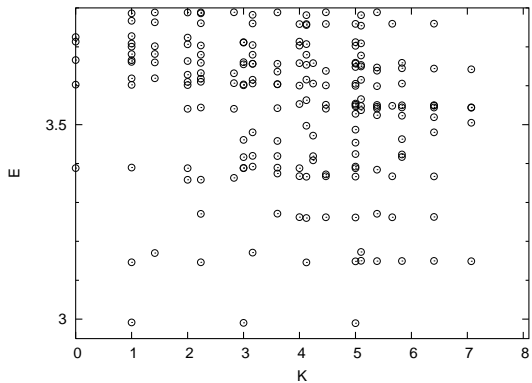}}
\caption{\label{fig18} Energy spectrum of the $V_2$ model
for $N=10$ at $a/b=0.35$. The ground state at $\mathbf{K}=(0,5)$ and 
states at $\mathbf{K}=(0,3),(0,1)$ are quasidegenerate.}
\end{center}
\end{figure}
Fig. \ref{fig18} shows the energy spectrum at $a/b=0.35$. The ground state
at $\mathbf{K}=(0,5)$ and the lowest energy states at 
$\mathbf{K}=(0,3),(0,1)$ are quasidegenerate.
In Fig. \ref{fig18},  clearly separated low-lying bands are seen, 
which is consistent with the existence of particle-hole excitations 
 of a stripe state.
The direction of stripes is always parallel to the $x$-direction.

\begin{figure}[ht]
\rotatebox{-90}{
\includegraphics[width=5cm]{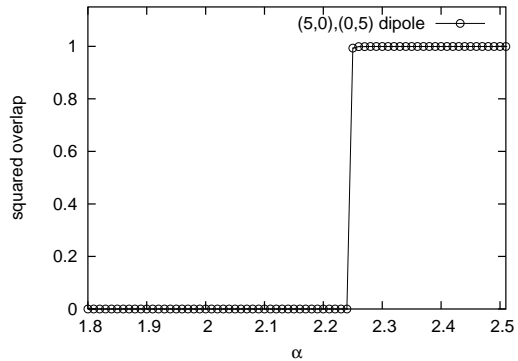}}
\caption{\label{fig19} The squared overlaps of
the two degenerate ground states
of the dipolar model
with the ground states of  the $V_2$ model
are shown for $N=10$ at $a/b=1.0$. 
We present results for 
the two ground states at $\mathbf K=(5,0),(0,5)$.
$\mathbf{K}=(0,5)$ and $\mathbf{K}=(5,0)$ are exactly degenerate 
due to geometrical symmetry at $a/b=1.0$.}
\end{figure}

Fig.\ref{fig19} shows the overlap of the ground state of the dipolar 
model with the ground state of the $V_2$ model. 
The overlap is close to 1.0 for $\alpha$ beyond $\sim 2.2$. 
In the pseudopotential expansion, the dipolar model  has 
higher order terms beyond $V_2$ but this result suggests that, 
for aspect ratios above $\sim 0.5$,   quasidegenerate ground states 
for the dipolar model at large $\alpha$ coincide with those of the $V_2$ model.

\section{\label{sec:level4} RESULTS FOR the three-body interaction}

\begin{figure}
\rotatebox{-90}{
\includegraphics[width=5cm]{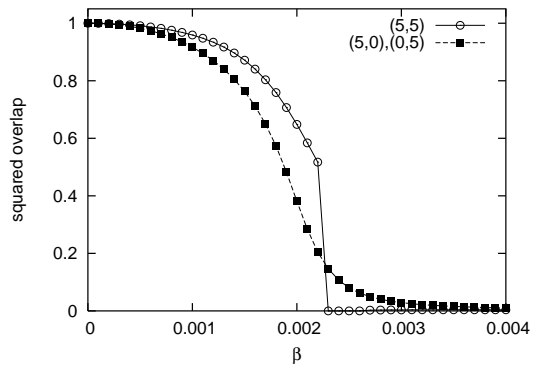}}
\caption{\label{fig20} The squared overlaps of
the two degenerate ground states and a nearly degenerate state
for the $V_{\mathrm{3b}}+V_2$ model
with the Pfaffian states are shown for $N=10$ at $a/b=1.0$. 
The Pfaffian states exist at $\mathbf K=(5,5),(5,0),(0,5)$.
The two ground states of the $V_{\mathrm{3b}}+V_2$ model
 appear at $\mathbf K=(5,0),(0,5)$.
$\mathbf{K}=(0,5)$ and $\mathbf{K}=(5,0)$ are exactly degenerate 
due to geometrical symmetry at $a/b=1.0$.
}
\end{figure}

In this section, we present results for the 
$V_{\mathrm{3b}}+V_2$ model.  We set  $\beta \equiv V_2/C_{3b}$.
We add a moderate $V_2$ interaction (small $\beta$) and examine the stability
of the Pfaffian states which are the exact zero-energy ground states at $\beta=0$. 
In Fig. \ref{fig20}, we show the squared overlaps of nearly degenerate 
ground states
 at $a/b=1.0$ with the Pfaffian states at each $\mathbf{K}$  where Pfaffian states exist.  
The overlaps decrease monotonically as $\beta$ increases. 
The overlaps drop around $\beta \sim 2.0 \times 10^{-3}$. 
The collapse is not as sharp as that of the dipolar model. 
For other aspect ratios we have investigated, this collapse always occurs. 
The value of $\beta$ where this collapse occurs increases as $a/b$ decreases.
\begin{figure}
\begin{center}
\begin{tabular}{c}
\rotatebox{-90}{
\includegraphics[width=5cm]{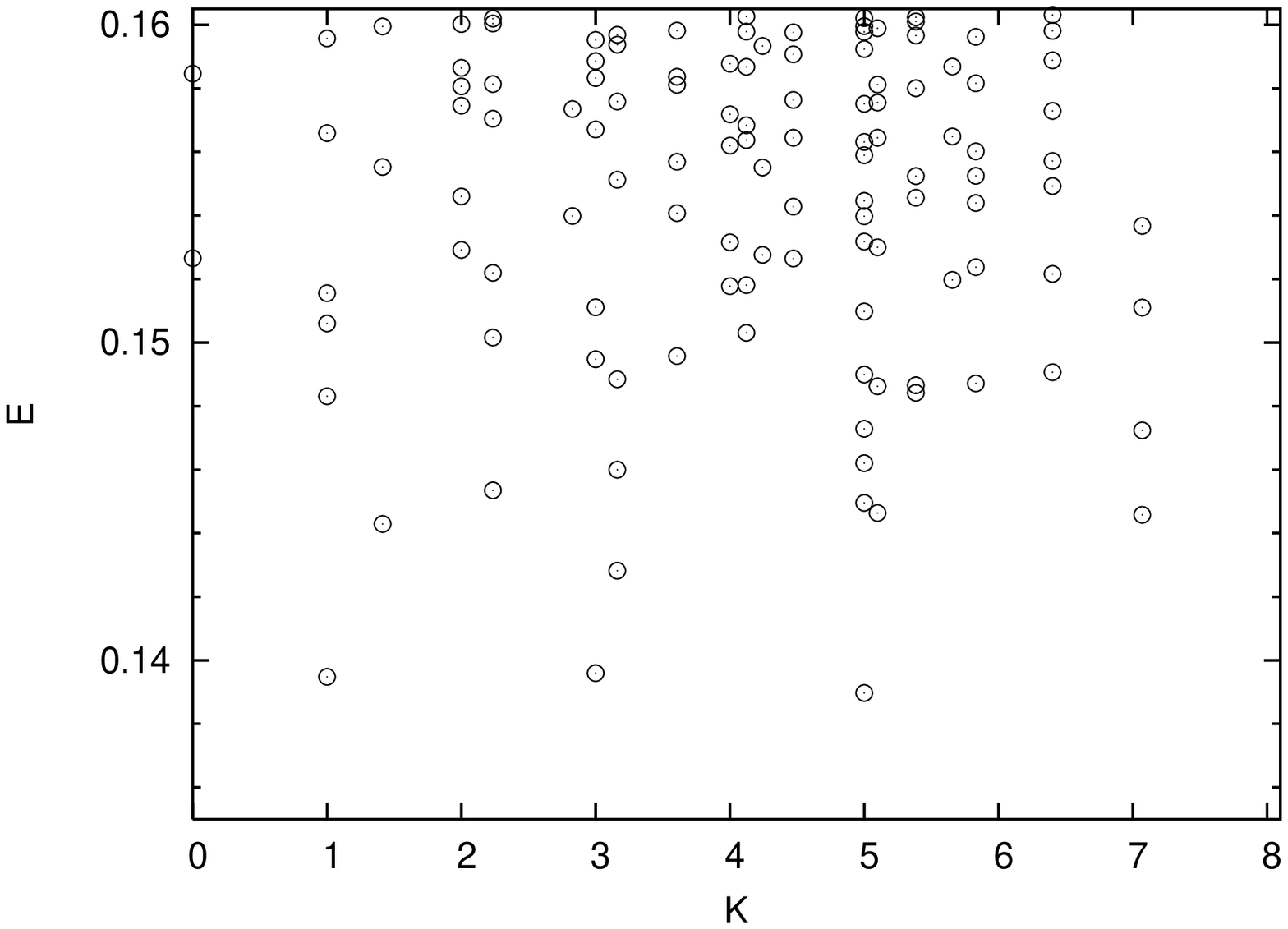}} \\
\end{tabular}
\caption{\label{fig21} Energy spectrum of the 
$V_{\mathrm{3b}}+V_2$ model for $N=10$ at $\beta=1.0 \times 10^{-2}, a/b=0.8$. 
The ground state at $\mathbf{K}=(0,5)$ and 
states at $\mathbf{K}=(0,1),(0,3)$ are quasidegenerate.}
\end{center}
\end{figure}
\begin{figure}
\begin{center}
\rotatebox{-90}{
\includegraphics[width=5cm]{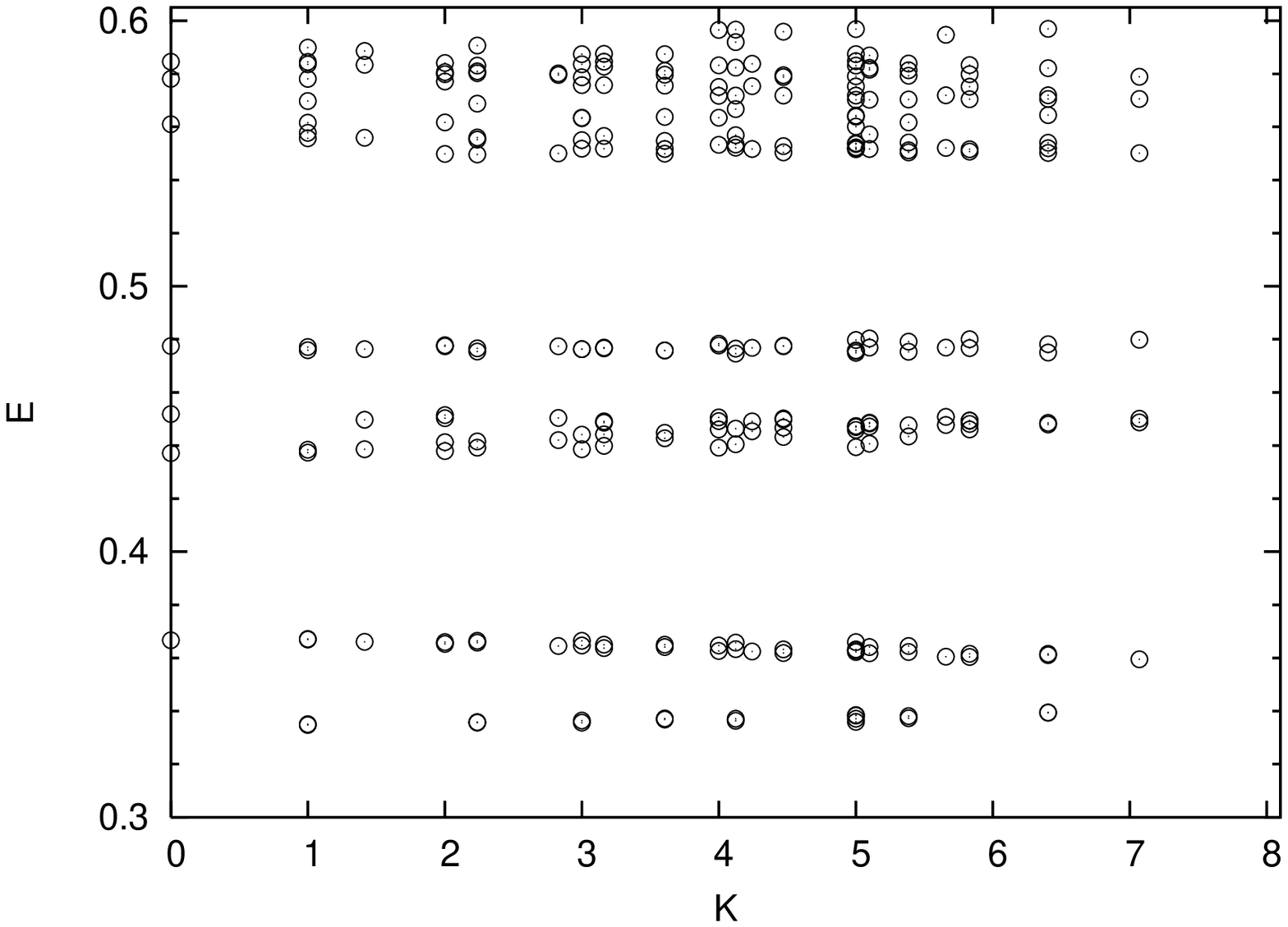}}
\caption{\label{fig22} Energy spectrum of the 
$V_{\mathrm{3b}}+V_2$ model for $N=10$ at $\beta=0.1,a/b=0.8$. 
The quasidegenerate states are 
at $\mathbf{K}=$(1,0),(0,1),(2,1),(1,2)(3,0),(0,3),(2,3),(3,2)
,(4,1),(1,4),(4,3),(3,4),(5,0),(0,5),(4,5),(5,4),(2,5),(5,2).
}
\end{center}
\end{figure}

Let us next turn to states after the collapse. 
At high aspect ratios, we observe quasidegenerate ground states 
forming stripes.
The stripe state changes to a bubble state around 
$\beta \sim 2.5 \times 10^{-2} $. 
As we further increase $\beta$, 
the ground state starts to have a large overlap with the ground
state of the $V_2$ model at $\beta$ above $1.0$.
The structure of the energy spectrum is similar to that of the $V_2$
model. 

As an example of high aspect ratios, 
we present the energy spectrum for $N=10$ at 
$a/b=0.8,\beta=1.0 \times 10^{-2}$ in Fig. \ref{fig21}. The ground state at $\mathbf{K}=(0,5)$
and states at $\mathbf{K}=(0,3),(0,1)$ are quasidegenerate. These states
form a stripe state.
The direction of stripes is parallel to the $x$-direction and the number of 
stripes is two while
 first-excited levels are not clearly separated in Fig. \ref{fig21}.
The pair distribution function $G(\mathbf{r})$ for the ground state 
at $\mathbf{K}=(0,5)$ is similar to that of the stripe state of the 
dipolar model  presented in Fig. \ref{fig4}.
When we decrease the aspect ratio, the stripe state changes to an intermediate
state  at $a/b \sim 0.5$. If we further decrease the aspect ratio, a stripe 
state reappear at $ a/b$ below $ 0.3$.

A bubble state appears at high aspect ratios for $\beta =0.025 \sim 1.0$.
In Fig. \ref{fig22}, we present the energy spectrum for 
$N=10$ at $ \beta=0.1,a/b=0.8$. 
There are quasidegenerate ground states arrayed in the lattice 
with the primitive vectors $\mathbf{e}_1$ and $\mathbf{e}_2$, 
indicating a formation of bubbles. 
The presence of  clearly separated  bands 
in Fig. \ref{fig22} gives further evidence of it.
The $G(\mathbf{r})$ of quasidegenerate states 
is similar to that of the bubble state  of the 
dipolar model  presented in  Fig. \ref{fig8}.
At low aspect ratios $ a/b$ below $ 0.3$, 
the quasidegenerate ground states remain to form a stripe state. 
The direction of stripes is always parallel to the $x$-direction. 
The number of stripes varies for $N$ and $a/b$.
Intermediate states appear in the region $a/b = 0.3 \sim  0.5$.

\section{\label{sec:level6} Conclusions}
We have exhibited evidence that bosons 
with a dipole moment
in the lowest Landau level at $\nu=1$ have  
 phases of incompressible liquid, stripes and bubbles 
based on exact diagonalizations up to $N=12$.
While a moderate amount of the dipolar interaction stabilizes 
the incompressible liquid, a further amount induces a collapse of it.
The state after the collapse is a compressible state. 
Up to $N=12$, it is a stripe state or 
an intermediate state between stripes and bubbles. 
The number and the direction of stripes are sensitive to 
 the aspect ratio,  the strength of the interaction 
and the number of particles. 
As the dipolar interaction gets strong, two kinds of 
bubble states appear for high aspect ratios.
 The transition 
between them is observed as the dipolar interaction gets strong. 
On the other hand, for small aspect ratios, stripe states remain to appear 
even when  the dipolar interaction is rather strong.

We have also considered the $V_2$ model where 
the bosons do not have a hard-core.
It has turned out that it does not have a phase of incompressible liquid. 
Its ground state is a bubble state for aspect ratios above $0.5$. 
A certain amount of the $V_0$  interaction induces 
a transition to another bubble state a different array.
We have observed that, for high aspect ratios, 
the quasidegenerate ground states for the dipolar interaction 
at the strong limit 
have large overlaps with those of  the $V_2$ model.
For low aspect ratios  below 
$0.5$, quasidegenerate ground states of the $V_2$ model are identified 
 to be a stripe state. 

We have also studied  the model where the bosons interact through 
the three-body contact interaction for which the Pfaffian state is the exact ground state.  
It shows similar properties as those of  the dipolar model : our results  
give evidence that the model has  phases of stripes and bubbles.

Finally we remark that the geometry treated in this paper 
 is a subset of  the general geometry of torus where 
the period vectors are not necessarily perpendicular.

{\it Acknowledgment.} 
A part of numerical calculations has been performed at the supercomputing  facility 
of Institute for Solid State Physics, Univ.of Tokyo.
K.I. is partially supported by Grant-in-Aid 
for Science Promotion and 
also by the SAKURA project organized by JSPS and CNRS.

{\it Note Added.} While we were completing this work, 
a preprint which has a partial overlap with this paper
 appeared \cite{Chung07}. 
Energy levels which show  degenerate ground states which are characteristic of the Pfaffian and stripes are observed at a strength of the dipolar interaction.


\begin{thebibliography}{Lamport 90}
\bibitem{N=8.3} For $N=8$, a stripe state with three stripes does not appear.
\end{thebibliography}
\end{document}